\documentclass[manuscript]{aastex}

\usepackage{graphicx}
\usepackage{dcolumn}
\usepackage{bm}
\usepackage{amssymb}
\begin{document}


\title{A new estimator of the deceleration parameter from galaxy rotation curves}
\author{Maurice H.P.M. van Putten\footnote{Corresponding author. E-mail: mvp@sejong.ac.kr}} 
\affil{$^1$Sejong University, 98 Gunja-Dong Gwangin-gu, Seoul 143-747, Korea and
Kavli Institute for Theoretical Physics, University of California, Santa Barbara, CA 93106-4030 USA}	

\begin{abstract}
 The nature of dark energy may be probed by the derivative $Q=\left.dq(z)/dz\right|_0$ at redshift $z=0$ of the deceleration parameter $q(z)$.  It is probably static if $Q<1$ or dynamic if $Q>2.5$, supporting $\Lambda$CDM or, respectively, $\Lambda=(1-q)H^2$, where $H$ denotes the Hubble parameter. We derive $q=1-\left(4\pi a_0/cH\right)^{2}$, enabling a determination of $q(z)$ by measurement of Milgrom's parameter $a_0(z)$ in galaxy rotation curves, equivalent to the coefficient $A$ in the Tully-Fisher relation $V^4_c=AM_b$ between rotation velocity $V_c$ and baryonic mass $M_b$. We infer that dark matter should be extremely light with clustering limited to the size of galaxy clusters. The associated transition radius to non-Newtonian gravity may conceivably be probed in a free fall Cavendish type experiment in space.
\end{abstract}

\nopagebreak

\maketitle

\section{Introduction}

Large scale cosmology is to leading order described by a Friedmann-Robertson-Walker line-element
\begin{eqnarray}
ds^2  = - dt^2 + a(t)^2 \left(dx^2 + dy^2 + dz^2\right)
\label{EQN_FRW}
\end{eqnarray} 
with dynamical scale factor $a(t)$. Here,
the evolution of $a(t)$ is paramaterized by $H=\dot{a}/{a}$ and the deceleration parameter $q=-H^{-2}\ddot{a}/a$, where the dot refers to differentiation with respect to time. In evolving (\ref{EQN_FRW}) by general relativity, a dark energy density $\Lambda/8\pi>0$ is inferred from the observed three-flat cosmology with deceleration
\begin{eqnarray}
q=\frac{1}{2}\Omega_M-\Omega_\Lambda<0
\label{EQN_E1}
\end{eqnarray}
in Type Ia supernova surveys \citep{rie98,per99}. Here, $\Omega_M=\rho_M/\rho_c$ and $\Omega_\Lambda=\Lambda/8\pi \rho_c$, where $\rho_c=3H^2/8\pi$ denotes the closure density and $\Lambda$ is commonly referred to as a cosmological constant. The value $\Omega_\Lambda\simeq 0.7$ suggests that our cosmology is presently approaching a de Sitter state with a cosmological horizon at the Hubble radius $R=c/H_0$. A de Sitter state is fully Lorentz invariant, in our current Universe broken only by the presence of a minor amount of matter.

A telltale signature of dark energy in accelerated cosmological expansion (\ref{EQN_E1}) is its static or dynamic behavior. Here, we consider the problem of discriminating between $\Lambda$CDM and a dynamic dark energy in the form of $\Lambda=(1-q)H^2$, recently proposed as a back reaction of thermodynamic properties of the cosmological horizon \citep{van15b} motivated by holographic arguments \citep{bek81,tho93,sus95,van12} and a modified Gibbons-Hawking temperature \citep{unr76,gib77,cai05}. This dynamical dark energy has the property that it vanishies in the radiation dominated era, leaving baryon nucleosynthesis unaffected. In this era, the surface gravity of the cosmological horizon vanishes, when it touches the light cone of distant inertial observers. These two alternatives predict distinct values of the derivative $Q=\left.dq(z)/dz\right.$ at redshift $z=0$:
\begin{eqnarray}
Q_{stat}<1,~~~Q_{dyn}>2.5
\label{EQN_Q}
\end{eqnarray}
in $\Lambda$CDM and, respectively, $\Lambda=(1-q)H^2$. Illustrative for a holographic origin of $\Lambda$ is a dimensional analysis based on $L_0=c^5/G$ and the associated pressure $p=L_0c/A_H$ on the cosmological horizon, where $A_H=4\pi R_H^2$. In a pure de Sitter space $(q=-1)$, $\rho_\Lambda=-p$ by Lorentz invariance, whereby $\Omega_\Lambda=2/3$ in remarkable agreement with observations. See also \citep{eas11} for a derivation based on entropic forces. 

By (\ref{EQN_Q}), $q(z)$ can be used to distinguish between $\Lambda$CDM and a dynamical $\Lambda$, provided it is resolved sufficiently accurately about $z=0$. Current data from Type Ia supernova surveys, however, seem inconclusive, that appears to be due to systematic errors possibly related to the tension with Planck data on the Hubble parameter \citep{pla13,pla14}.

Here, we consider a new probe of $q(z)$ in galaxy rotation curves and its implications for clustering of dark matter. This approach is based on a finite sensitivity of weak gravity to $\Lambda$ (static or dynamic). Gravitational attraction beyond what is inferred from (luminous) baryonic matter is generally observed in galaxies and galaxy clusters \citep{fam12} at accelerations of 1\,\AA\,s$^{-2}$ or less. This apparent non-Newtonian behavior is commonly attributed to dark matter, based on the success of Newton's theory of gravity in the solar system and its extension to strong gravity by embedding in general relativity. Supporting data for the latter derive from orbital motions at accelerations $a=R_g(c/r)^2\simeq  10^{-6}-10^{-2}$ m\,s$^{-2}$ of planets in the solar system at distances $r$, where $R_g = GM_\odot/c^2\simeq 1.5$ km denotes the gravitational radius of the Sun with Newton's constant $G$. Its extension in general relativity to higher accelerations has been fully vindicated by precession measurements in the Hulse-Taylor binary pulsar PSR1913+16 ($a=10^0-10^2$ m\,s$^{-2}$) \citep{hul75}. However, our observations of dark matter take us to the opposite limit of extremely weak gravity, not probed by our solar system or strong field counterparts in compact binaries. The parameter regime of about 1\,\AA\,s$^{-2}$ takes us away from existing tests of Newtonian gravity by a factor of about $10^4$, which is not small. Importantly, this scale is similar to the scale of cosmological acceleration $a_H=cH$, where $c$ denotes the velocity of light and $H$ is the Hubble parameter. Currently, $H_0 \simeq 67$ km s$^{-1}$Mpc$^{-1}$ \citep{pla13,pla14}.

To realize our new probe of $q(z)$, we consider weak gravity on the cosmological background (\ref{EQN_FRW}) parameterized by $(H,q)$ in a recent formulation of unitary holography
\citep{van15a}. In geometrical units with Newton's constant $G$ and the velocity of light $c$ equal to unity, $\Lambda$ and $\sqrt{\Lambda}$ are of dimension cm$^{-2}$ and, respectively, cm$^{-1}$, corresponding to dark energy volume and, respectively, surface density. The latter may be recognized as the thermal energy density $\Sigma = \frac{1}{2}T= {H}/(4\pi)$, defined by a de Sitter temperature $T_{dS}=H/2\pi$ \citep{gib77}. While $\Lambda/8\pi$ is notoriously small, $\Lambda \simeq 1.21 \times 10^{-56} \,\mbox{cm}^{-2}$, $\Sigma= \sqrt{\Lambda}/(4\sqrt{2}\pi) \simeq 6\times 10^{-29} \,\mbox{cm}^{-28}$ is not. An immediate implication is a critical transition radius for gravitational attraction. Around a central mass $M=M_{11}10^{11}M_\odot$ of a typical galaxy, $A\Sigma = M$ with for a two-sphere with area $A=4\pi r^2$, giving a transition radius
\begin{eqnarray}
r_t = \sqrt{MR_H} = 4.6 \,M_{11}^\frac{1}{2} \, \mbox{kpc}.
\label{EQN_rt}
\end{eqnarray}
The transition radius (\ref{EQN_rt}) is common to galaxy rotation curves and bears out in well in $a_t/a_H\simeq 0.1$ $(a_t = GM/r_t^2)$ in a deviation of centripital accelerations $a$ relative to the Newtonian acceleration $a_N$ expected from the observed baryonic mass \citep{mil83,fam12}, here shown in Fig. 1. (\ref{EQN_rt}) defines {\em strong} gravitational interactions in $r<<r_t$ and {\em weak} gravitational interactions in $r>>r_t$ with accelerations, respectively,
\begin{eqnarray}
a<<a_H,~~a>>a_H.
\end{eqnarray} 
In geometrical units, holography hereby identifies $a_H$ as a critical acceleration in galaxy rotation curves.

\begin{figure}[h]
	\centerline{\includegraphics[scale=0.55]{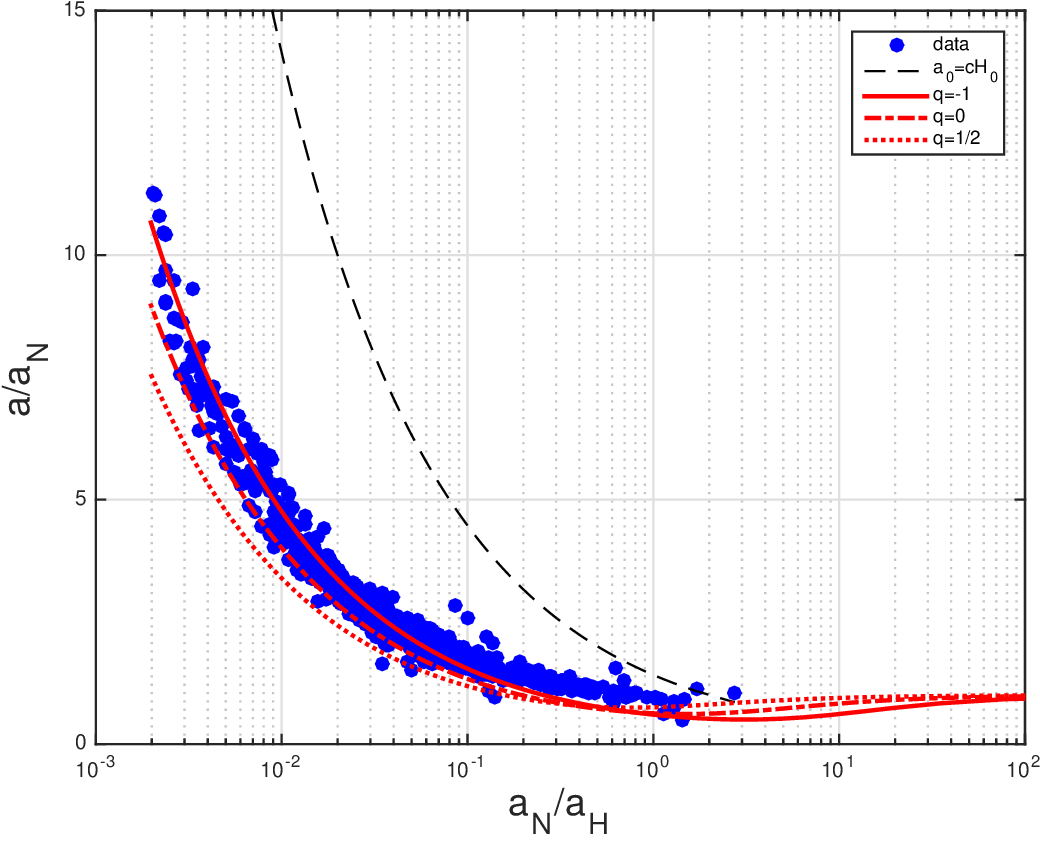}}
	\caption{Galaxy rotation curves (blue dots) reveal a transition to a $1/r$ force law at weak accelerations asymptotically in $a<<a_H$ away from Newtonian forces in $a>>a_H$ based on the observed baryonic matter. Shown is a theoretical curve (red) in unitary holography with a good match in a cosmological background with deceleration parameter close $q$ in the range $-1<q<-0.5$. Data are from galaxy curves with essentially zero redshifts from \cite{fam12}.}
\end{figure}

In \S2, we express gravitational attraction in terms of a conformal factor, encoding information on particle positions in unitary holography. The Newtonian limit is recovered in \S3, extended to non-Newtonian asymptotic behavior in $r>>r_t$ in (\ref{EQN_rt}) on a de Sitter background in \S4. A further extension to (\ref{EQN_FRW}) is given in \S5, wherein a finite sensitivity in $r>>r_t$ to $\Lambda$ is proposed as a new estimator for $q(z)$, proposed to determine (\ref{EQN_Q}).  We summarize our theory in \S6 with an outlook on future tests.

\section{Conformal factors from distance information}

Unitary holography expresses distances of particles of mass $m$ to time-like holographic two-surfaces in terms of information $I=2\pi \Delta \varphi$ defined by a total phase difference $\varphi=kr$ derived from its propagator with Compton wave number $k=mc/\hbar$, where $\hbar$ denotes the Planck constant. Holographic imaging is hereby an extension of holographic bounds originally developed for black hole spacetimes \citep{bek81,tho93,sus95} to spacetime outside black hole event horizons. Thus, $m$ is a holographic superposition of $A/l_p^2$ light modes determined by the hyperbolic structure of spacetime, where $l_p=\sqrt{G\hbar/c^3}$ denotes the Planck length. 

This approach has two consequences. First, on macroscopic scales, $A/l_p^2$ is astronomically large. The holographic modes are extremely light with an energy scale
\begin{eqnarray}
\epsilon = \frac{mc^2l_p^2}{A},
\label{EQN_eps}
\end{eqnarray}
that introduces a sensitivity to any similarly low energy scale in the background vacuum. The latter is described by the elliptic structure of spacetime, that governs gravitational attraction. (In general relativity, the elliptic part embeds Newton's law of gravity in a conformal factor.) Second, holographic imaging is a function of $A\Omega$, where $A$ is the area of bounding surface and $\Omega$ is the projection opening angle of its surface elements. Factorization of $A\Omega$ is hereby an internal symmetry of holography \citep[cf.][]{tho15}. Scaling of $A$ and $\Omega$ corresponds to curvature and, respectively, lensing. These may be realized by a conformal factor or a deficit angle, that are essentially different manifestations of the same. 

In encoding $I$ in $fA=A-A_E$ or, equivalently, $f\Omega = 4\pi - \Omega_E$, $A_E=8\pi ms$ and $\Omega_E=8\pi m/s$ are the Einstein area and opening angles, respectively, i.e.: $fA\Omega= 4\pi\left(A - A_E\right) = 4\pi A \left(4\pi - \Omega_E\right) = 16\pi^2s^2 f.$ Here, the factor of four in $A_E=4Il_p^2$ derives from a counting argument on the minimal number of bits required to encode matter and fields \citep{van15a}. A holographic screen hereby attains minimal size with $A=A_E$ or $\Omega_E=4\pi$ at the Schwarzschild radius $s=R_S$, $R_S=2m=\sqrt{S/\pi}$ with $S=\min I=4\pi m^2$ in $I = 2\pi m \left(s-R_S\right)+ S$ equal to the Bekenstein-Hawking entropy. Accordingly, we have 
\begin{eqnarray}
f=1-\frac{2m}{s}.
\end{eqnarray}

In general relativity, the gravitational field about a point mass can be described by a conformal factor $\Phi$ in an isotropic line-element
\begin{eqnarray}
ds^2 = - N^2 dt^2 + \Phi^4\left(dx^2 + dy^2 + dz^2\right),
\label{EQN_A1}
\end{eqnarray} 
where $N=N(\Phi)$ denotes the gravitational redshift, i.e., the ratio of energy-at-infinity to locally measured energy. According to the above, $R_S=\sqrt{4S/\pi}$ expresses the mass-energy of a particle by its linear size, locally measured by the minimal surface area $4S$ of an enveloping holographic screen. We are at liberty to choose a gauge 
\begin{eqnarray}
N\Phi^2\simeq\mbox{const.},
\label{EQN_G}
\end{eqnarray} 
defined by a constant total mass energy-at-infinity in the approximation of small perturbations to the spherically symmetric line-element (\ref{EQN_A1}). For a detailed consideration of such time-symmetric data, see \citep{van12}, where it serves as a condition in the application of Gibbs' principle in entropic forces in black hole binaries. According to the equations of geodesic motion, Newton's law then derives from $N$ in the large distance limit. With $dx^2+dy^2+dz^2= d\rho^2 + \rho^2d\theta^2 + \rho^2\sin^2\theta d\phi^2$ rexpressed in spherical coordinates $(\rho,\theta,\phi)$, $\rho$ reduces to the ordinary radial distance $r$ at large separations, (\ref{EQN_A1}-\ref{EQN_G}) embed Newton's law in
\begin{eqnarray}
\Phi \simeq f^{-\frac{1}{4}}\simeq 1 + \frac{m}{2r}~~(r>>2m).
\label{EQN_C}
\end{eqnarray}

\section{Newtonian limit in ordinary vacuum}

In what follows, unless otherwise specified, $A$ shall refer to surface area as well as the number of Planck sized surface elements $A/l_p^2$.

In holography, the wave function of a particle $m$ results from $A$ Planck sized harmonic oscillators of low energy (\ref{EQN_eps}). Ordinarily, one mode in the image appears for each mode in the screen. (The dimension of the phase space in the image equals the number of degrees of freedom in the screen.) Quantum mechanically, $m$ is the time rate-of-change of total phase as measured at infinity,
\begin{eqnarray}
m=\frac{1}{2}A\omega
\label{EQN_m}
\end{eqnarray}
of the ground state energies $(1/2)\omega$ of each harmonic oscillator in the screen. Distance encoding derives from aforementioned $\Delta\varphi = kr$ with the total wave number $k$ given by the superposition of these massless modes,
\begin{eqnarray}
k=\frac{1}{2}A\kappa,
\label{EQN_k}
\end{eqnarray} 
based on the trivial dispersion relation $\kappa=\omega$ of ordinary vacuum, that recovers the Compton wave number $k=k_C$, $k_C=m$, with low energy frequencies
\begin{eqnarray}
\omega_N = \frac{2m}{A}=\frac{a_N}{2\pi}
\label{EQN_kN}
\end{eqnarray}
defined by the Newtonian acceleration $a_N={m}/{r^2}.$

The Compton relation $k=m$ recovered by the trivial dispersion $\kappa_N=\omega_N$ associates $\kappa_N$ with the Unruh temperature of Newtonian acceleration (\ref{EQN_kN}). 

In entropic gravity \citep{ver11}, the above implies entropic forces on a test particle of mass $m'$ at screen temperatures $T=m/2\pi r^2$ by $dS=-dI = - 2\pi m' dr$, giving $F=-dU/dr=TdS/dr=-mm'/r^2$. In keeping with (\ref{EQN_G}-\ref{EQN_C}), however, we shall not pursue these arguments here.

\section{Sensitivity to $\sqrt{\Lambda}$ in a de Sitter background}

By (\ref{EQN_eps}), (\ref{EQN_kN}) is susceptable to low energy de Sitter temperatures of the cosmological horizon. Screen modes satisfy the dispersion relation 
\begin{eqnarray}
\omega=\sqrt{\kappa^2+\omega_H^2}
\label{EQN_DR1}
\end{eqnarray}
representing an incoherent sum of a momenta $\kappa$ and the background de Sitter temperature, $\omega_H=T_{dS}$ \citep{nar96,des97,jac98}. A spherical screen imaging a mass $m$ at its center hereby assumes
\begin{eqnarray}
\omega_N=\omega-\kappa_H:~~\kappa=\sqrt{\omega_N^2+2\omega_N\omega_H},
\label{EQN_T1}
\end{eqnarray}
giving $\kappa\simeq \omega_N$ $(r>>r_t$) and $\kappa\simeq \sqrt{a_0a_N}$ $(r<<r_t)$ with $a_0=2a_H$ as proposed in \citep{kli11}. However, $\kappa(\omega_N)$ from (\ref{EQN_T1}) overestimates the Milgrom parameter $a_0$ \citep{mil83} by about {\em one order of magnitude} according to the data shown in (Fig. 1). Here, Milgrom's parameter is equivalent to the coefficient $A$ in the Tully-Fisher relation $V^4_c=AM_b$, where $V_c$ denotes the rotation velocity in a galaxy of baryonic mass $M_b$ \citep{mcg11a,mcg11b}. The wave number $\kappa_N$ in (\ref{EQN_T1}) is {\em not} representative for $\kappa$ of the image within.

On a background (\ref{EQN_FRW}) with $\Lambda>0$, image modes satisfy the dispersion relation 
\begin{eqnarray} 
\omega^\prime = \sqrt{\kappa^2 + \Lambda}
\label{EQN_DR2}
\end{eqnarray}
defined by the wave equation of a vector field in curved spacetime by coupling to the Ricci tensor $R_{ab}=\Lambda g_{ab}$. This applies to the electromagnetic vector potential \citep[e.g.][]{wal84} as well as the Riemann-Cartan connections in SO(3,1) in a Lorenz gauge \citep{van96}. It implies an effective rest mass energy $\sqrt{\Lambda}$ of the photon and graviton and photon. Effective mass is not the same as true mass. Even so, we mention in passing that the problem of consistent general relativity with massive gravitons has recently received considerable attention \citep{der11,ber14}. With $−q_0 H^2= H^2 + \dot{H}$ , the generalized Higuchi constraint $m^2 \ge 2(H^2 +\dot{H})$ \citep{hig87,des01,gri10} reduces to $\Omega_\Lambda\ge − 2 q_0$. Based on observations, $−1 < q_0 < −0.5$ \citep{rie04,wu08,gio12}, whereby $q_0 > -1$ appears secure at any rate. 

By (\ref{EQN_DR1}) and (\ref{EQN_DR2}), distinct effective masses appear in the kinetic energies $E=\omega-\omega_H$ and $E^\prime=\omega^\prime - \sqrt{\Lambda}$ of low energy modes in the screen and image, namely
\begin{eqnarray}
E \simeq \frac{\kappa^2}{2\kappa_H},~~E^\prime \simeq \frac{\kappa^2}{2\sqrt{\Lambda}}
\label{EQN_O}
\end{eqnarray}
($\kappa<<\kappa_H,\sqrt{\Lambda}$). In weak gravitation in de Sitter space, therefore, {\em a direct correspondence between screen and image modes is lost}, in striking departure from the above Newtonian limit in $r<<r_t$ in the previous section.

Specifically, (\ref{EQN_O}) shows a discrepancy by a factor of $2\sqrt{2}\pi$ in effective mass $\sqrt{\Lambda}$ over that in $\kappa_H$. A given $\kappa=\kappa(\omega_N)$ 
of screen modes has an associated reduced energy in the relatively more heavy image modes, satisfying
\begin{eqnarray}
\omega_N^\prime=\omega^\prime-\sqrt{\Lambda}=\sqrt{\kappa^2+\Lambda}-\sqrt{\Lambda}
\end{eqnarray}
with corresponding reduced screen momenta $\kappa^\prime = \sqrt{\omega_N^{\prime2}+2\omega_N^\prime\omega_H}.$
Fig. 1 shows the graph $\kappa^\prime(\omega_N)$ to be in agreement with the data. Specifically, we arrive at Milgrom's constant
\begin{eqnarray}
a_0 =\left( \frac{\kappa_H}{\sqrt{\Lambda}}\right)2cH_0 = \frac{cH_0}{\sqrt{2}\pi} \simeq 1.5\times 10^{-8} \,\mbox{cm}\,\mbox{s}^{-2},
\label{EQN_a0}
\end{eqnarray}
where we restored dimensions in cgs units.

\section{Sensitivity to $q(z)$ in a FRW background}

The above generalizes to general Friedmann-Robertson-Walker (FRW) universes with modified de Sitter temperature \citep{cai05,van15b}
\begin{eqnarray}
T_{dS}= \frac{1-q}{2}\frac{H}{2\pi}.
\label{EQN_TdS}
\end{eqnarray}
\begin{figure}[h]
\centerline{\includegraphics[scale=0.55]{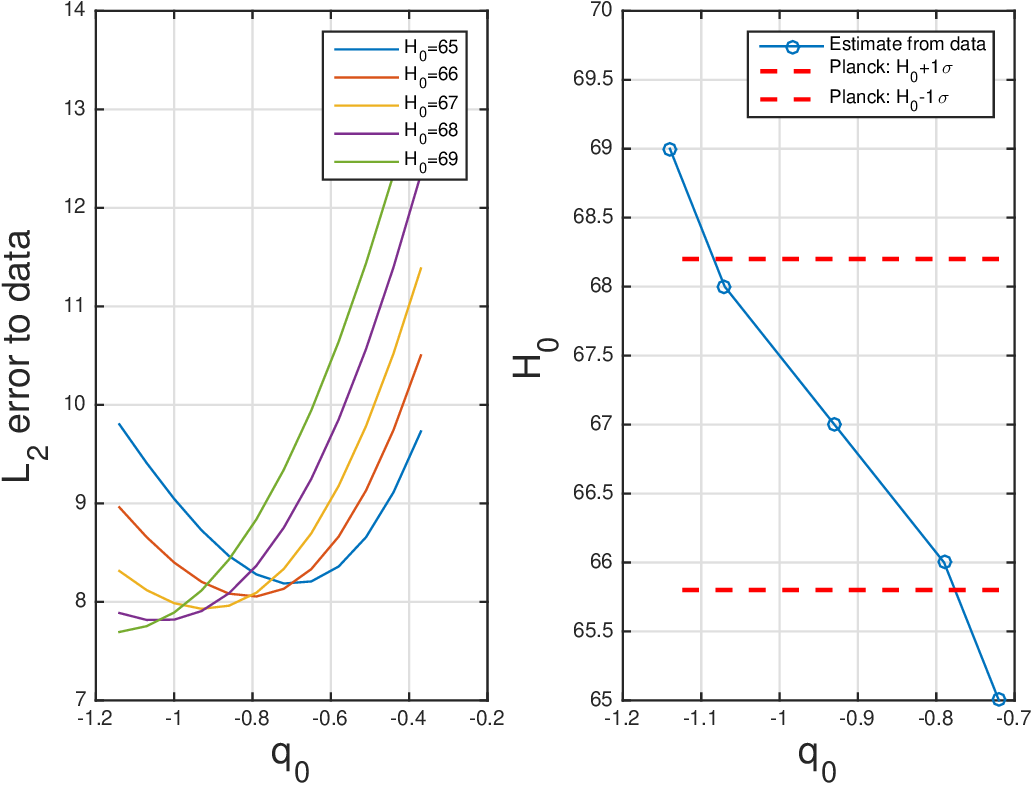}}
\caption{Estimation of $q_0$ by a least-squares fit to the \cite{fam12} sample of low redshift galaxies shown in Fig. 1 following rescaling to various $H_0$ in units of km s$^{-1}$ Mpc$^{-1}$. The resulting correlation $(q_0,H_0)$ is agrees with Planck data on a relatively low Hubble parameter of about 67 km s$^{-1}$ Mpc$^{-1}$.}
\end{figure}
A key feature of (\ref{EQN_TdS}) is that $T_{dS}=0$ in the radiation dominated era $q=1$, whereby it pertains only to relatively late time cosmologies, satisfying
\begin{eqnarray}
\Omega_\Lambda = \frac{1}{3}(1-q),~~\Omega_{CDM}=\frac{1}{3}(2+q).
\end{eqnarray}

As a consequence of (\ref{EQN_TdS}), Milgrom's constant attains the explicit expression
\begin{eqnarray}
a_0 = \frac{\sqrt{1-q}}{4\pi}cH,
\end{eqnarray}
allowing measurement of $q$ from $a_0$ as a function of redshift:
\begin{eqnarray}
q(z) = 1 - \left( \frac{4\pi a_0(z)}{cH(z)}\right)^{2}.
\label{EQN_qa0}
\end{eqnarray}

The existing low redshift sample of galaxies of \citep{fam12} recovers the value $-1<q_0<-0.8$ for the Planck estimate of $H_0$ (Fig. 2) and is broadly consistent with Type Ia supernova surveys \citep{rie04,wu08,gio12}.

More detailed future observations of $a_0(z)$ about $0\le z <<1$ offer a new venue
for determine
\begin{eqnarray}
Q=2(1-q_0)^2-2(1-q_0)a_0^{-1}\left.\frac{da_0(z)}{dz}\right|_{z=0}
\label{EQN_dqda}
\end{eqnarray} 
from a sample of galaxy rotation curves covering a finite range of low redshift $0\le z << 1$.

In \citep{van15b}, we considered the problem of discriminating between a dynamical and static $\Lambda$ parameter $(1-q)H^2$ versus $\Lambda$CDM. In the range of 
$-1<q_0<-0.6$, Fig. 2 in \citep{van15b} shows the disjoint ranges
\begin{eqnarray}
 0<Q_{stat}<1,~~2.5<Q_{dyn}\simeq 2.8
\label{EQN_R1}
\end{eqnarray}
associated with, respectively, $\Lambda$CDM and $\Lambda=(1-q)H^2$.
Approximating the general trend in the first with $dq(z)/dz \simeq (5/3)(1+q_0)$
and the second with $dq(z)/dz\simeq 2.7$, Fig. 3 shows the correlation
(\ref{EQN_dqda}) with $a_0^{-1}da_0(z)/dz$ at $z=0$.
\begin{figure}[h]
	\centerline{\includegraphics[scale=0.55]{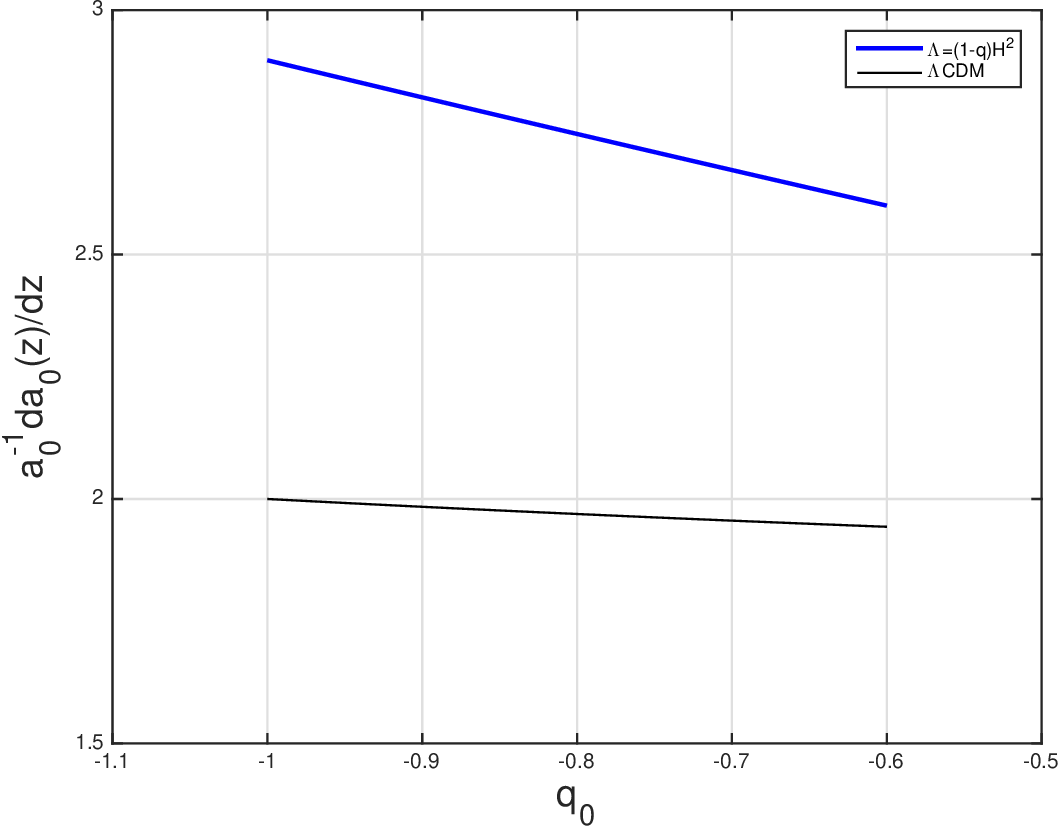}}
	\caption{Shown are the correlations of $a_0^{-1}da_0(z)/dz$ with $q_0$ for a dynamical dark energy $\Lambda(1-q)^2H^2$ and a static dark energy in $\Lambda$CDM.
	In $-1<q_0<-0.6$, these correlations are sufficiently distinct to be discriminated observationally, provided that $a_0(z)$ is measured accurately over a small range of low redshifts.}
\end{figure}

\section{Conclusions}

In unitary holography of matter, conformal factors encoding positions and gravitation attraction have a hidden low energy scale (\ref{EQN_eps}), that introduces a finite sensitivity to low energy scales in the cosmological background (\ref{EQN_FRW}) parameterized by $(H,q)$. This sensitivity is manifest in a transition to non-Newtonian gravitational attraction, that scales with inverse distance beyond a critical radius $r_t$ at accelerations on the scale of the surface gravity of the cosmological horizon. It produces Milgrom's law with a specific expression for the $a_0$ as function of $(H,q)$. This $a_0$ sensitivity to $(H,q)$ may be probed observationally in low redshift galaxy rotation curves.

By agreeent with data shown in Fig. 1, there is no apparent need for clustering of dark matter on the scale of galactic disks. Even so, there exists a cosmological distribution of dark matter \citep{van15b}. A major conclusion of the present work, therefore, is that dark matter must be extremely light, giving clustering on the scales of galaxy clusters but not down to the much smaller scales of galaxies. Conceivably, the putative dark matter particle is the lightest element in the Universe and may not be readily detectable in a laboratory experiment based on interactions with ordinary matter. 

We propose probing the static or dynamic nature of dark energy by $dq(z)/dz$. Values less than 1 or greater than 2.5 supporting $\Lambda$CDM, respectively, $\Lambda=(1-q)H^2$, here formulated in terms of $a_0^{-1}da_0/dz$ less than 2, respectively, greater than 2.5. These data may be obtained from an extended sample of low redshift galaxy rotation curves. 

Finally, scaling of the transition radius (\ref{EQN_rt}) to laboratory test masses, $r_t \simeq 1\,\mbox{cm}\,M_0^\frac{1}{2}$ with $M=M_0$\,g suggests a possible laboratory test, probing the proposed sensitivity to the cosmological background by a space based free fall Cavendish experiment.

{\bf Acknowledgments.} The author gratefully acknowledges stimulating discussions with G. 't Hooft, D.M. Eardley, S.S. McGaugh and constructive comments from the referee. The author thanks the Kavli Instiute for Theoretical Physics, UCSB, were some of the work has been performed. This report NSF-KITP-16-016 was supported in part by the National Research Foundation of Korea and the National Science Foundation under Grant No. NSF PHY11-25915.

\end{document}